\documentclass[conference]{IEEEtran}

\usepackage{cite}
\usepackage{amsmath}
\usepackage{mwe}
\usepackage{subfig}
\usepackage{setspace}
\usepackage{algorithm}
\usepackage{algpseudocode}
\usepackage{graphicx,pstricks}
\usepackage{float}
\usepackage{amsthm}
\usepackage{commath}
\usepackage{ifpdf}
\usepackage{cuted}
\usepackage{lipsum}
\usepackage{diagbox}
\usepackage{multirow}
\usepackage{romannum}
\usepackage{etoolbox}

\newtheorem{theorem}{Theorem}
\newtheorem{remark}{Remark}
\newtheorem{example}{Example}

\usepackage{amsfonts, amsmath, amssymb, epsfig, graphicx, wrapfig, color}

\begin{document}

\title{On Erasure Broadcast Channels with Hard Deadlines}
\author{%
\IEEEauthorblockN{Zohreh Ovaisi, Natasha Devroye,  Hulya Seferoglu,  Besma Smida, Daniela Tuninetti
}%
\IEEEauthorblockA{ECE Department, University of Illinois at Chicago \\ 
\tt \{zovais2, devroye, hulya, smida, danielat\}@uic.edu}%
}

\maketitle

\begin{abstract}
This paper considers packet scheduling over a broadcast channel with packet erasures to multiple receivers with different messages (multiple uni-cast) each with possibly different hard deadline constraints. A novel metric is  proposed and evaluated: the {\it global deadline outage probability}, which gives the probability that the hard communication deadline is not met for at least one of the receivers.  The cut-set upper bound is derived and a scheduling policy is proposed to determine which receiver's packets should be sent in each time slot. This policy is shown to be optimal among all scheduling policies, {\em i.e.,} it achieves all boundary points of cut-set upper bounds when the transmitter knows the erasure patterns for all the receivers ahead of making the scheduling decision. An expression for the  global deadline outage probability is obtained for two receivers and is plotted and interpreted for various system parameters. These plots are not Monte-Carlo simulations, and hence the obtained expression may be used in the design of future downlink broadcast networks. 
Future extensions to per-user deadline outage probabilities as well as to scenarios with causal knowledge of the channel states are briefly discussed.
\end{abstract}

\section{Introduction}
\label{sec:intro}

Recently, applications have emerged in which latency rather than data rate alone, are of prime importance. In particular, there exists a latency, rate and reliability tradeoff for data communications, which, in information theory, have mainly been looked at in the asymptotic sense of reliability (using probability of error as a proxy) $\rightarrow 1$, and latency (using the number of channel uses or blocklength as a proxy) $\rightarrow \infty$. In this work, we seek to characterize a different tradeoff --  what (reliability, rate) pairs can be achieved for a given latency.

As downlink communications are of critical interest in the last hop of wireless systems, we formulate a packet scheduling problem with hard deadline constraints over a broadcast channel with packet erasures. In particular, a base station serves a number ``impatient receivers'' with hard deadline constraints, where periodically generated packets at the base station are expected to be delivered to each receiver before their respective hard deadlines. In this setup, we define a novel metric: 
the {\it global deadline outage probability} as the probability that the hard communication deadlines are not met for at least one of the receivers. Our goal is to develop a packet scheduling policy that minimizes the global outage probability.

{\bf Prior Work.}
The packet scheduling problem with hard deadline guarantees has been considered in past literature. Here, we mention those that are most relevant to our work.

The work most closely related to our formulation is that of Hou ~\cite{hou2009admission, hou2014scheduling}.  In ~\cite{hou2009admission} clients  transmit their randomly generated bits to an access point under hard deadline constraints over an erasure channel and receive ACK/NACK information through feedback. The channel state is assumed to be static, the deadline constraints are the same for all clients, the number of packets of each client is at most one per frame, all generated  within the same period. 
In both ~\cite{hou2014scheduling} and ~\cite{hou2009admission}, each client $n$ needs a long-term average throughput requirement of $q_n$  delivered jobs per interval. The author first analytically derived a condition for a set of clients to be feasible: a feasible region of a system is defined to be the set of all feasible $[q_n]$. Such region is characterized given a deadlines and link reliability of each user. He then proposed feasibility optimal scheduling policies, meaning that they can fulfill every feasible set of clients. (A set of clients are feasible if they fall into the feasible region and are said to be fulfilled under a policy, if the long term average throughput of each client $n$ is at least $q_n$ jobs per interval). We note that strict optimality is only shown for equal deadlines; heuristics are proposed for unequal deadlines. 

The problem examined here considers downlink transmission, multiple packets rather than one, and different deadlines for the different users. We will also assume -- in order to obtain an upper bound on real-world performance -- that the channel erasures  are all known a priori, making ACK/NACK information irrelevant. We propose an optimal scheduling policy that minimizes our newly defined metric, the global deadline outage probability.
Note that in  ~\cite{hou2009admission, hou2014scheduling}, the average throughput requirement (\% of data that must be delivered by the deadline) is defined individually for each user and each user tolerates not receiving a percentage of its packets. However, in this paper, a unique global deadline outage probability is defined for all users and is stricter in the sense that all users should receive 100\% of their packets, else the system is considered to be in outage.

Problems of a similar flavor to this one have been considered in the {\it multi-cast} setting in ~\cite{li2010throughput, tran2012adaptive,  fu2014dynamic, kim2014scheduling} in which all users want the {\it same} content. What we consider here is the notably more challenging {\it multiple uni-cast} setting, where all users want different information. In contrast to some of the work in ~\cite{li2010throughput, tran2012adaptive,  fu2014dynamic, kim2014scheduling}, we seek  to obtain more analytical insight through explicit characterization of the {\it deadline outage probability}, a new metric. 

If one prohibits the possibility of a packet being erased, a similar type of problem is considered in \cite{liu2016spatial}, where 
a routing and scheduling mechanism is developed over multi-hop wireless networks with deterministically-generated packets. Packet routing and scheduling decisions are made by taking into account different hard deadline constraints. 
As mentioned before, \cite{liu2016spatial} does not consider any packet losses (erasures) over links. We consider an erasure broadcast channel. 

If one is only interested in a single user (rather than the multi-user broadcast scenario considered here), in ~\cite{fu2006optimal} an optimal scheduling policy that maximizes average data throughput for a fixed amount of energy and deadline is developed in the presence of a point-to-point Gaussian fading channel with hard deadline constraints and three types of channel state information:  when the channel state is completely known (like here), when  only the current state is known just before the transmission, and when the channel state is not known at all.

Rate-control and scheduling problems when broadcasting {\em network coded packets} with similar hard deadline constraints over erasure channels are considered in~\cite{gangammanavar2010dynamic}, and a novel strategy that jointly determines the incoming flow rates and coding to maximize the weighted sum rate is developed by taking into account reliability requirements. Two linear encoding strategies that achieve the capacity and stability regions over an erasure broadcast channel with feedback are proposed in~\cite{sagduyu2013capacity}.
Since our formulation assumes the source has channel state information ahead of scheduling, coding is unnecessary and the main focus is on packet scheduling.

{\bf Contributions.}  The 
key contributions of this work are:

\noindent $\bullet$ In Section~\ref{sec:model} we present the system model, the problem formulation and propose a new metric, 
{\it the global deadline outage probability}, defined as the probability that the hard communication deadline is not met for at least one of the receivers.

\noindent $\bullet$ In Sections~\ref{sec:greedy}  and \ref{sec:global outage} we derive the information theoretic cut-set outer bound for each erasure pattern and propose a greedy scheduling policy to determine which receiver's packets should be sent in each time slot. This policy is shown to be optimal in the sense that it both achieves the cut-set bound for a given erasure pattern and also minimizes the global deadline outage probability when the  transmitter is assumed to have a Pretoria knowledge of the channel erasures for the coming block. We use this to plot the tradeoffs between latency (deadlines), reliability (global deadline outage probability) and rate (supported arrival rates), noting that the plots obtained are not obtained from Monte-Carlo simulation, but rather from our analytic expressions.

\noindent $\bullet$ In Section~\ref{sec:num} we propose heuristics for a more practical scenario where only the current (transmission) time slot is known to base station and also a scenario where neither the current time slot, nor any future time slots are known to the base station for similar deadline constraints.

\section{System Model and Problem Formulation} 
\label{sec:model}

In this section we describe the channel model and state the precise problem to be solved.
We consider a {\it discrete memoryless broadcast channel with erasures}. 
It consists of one transmitter (base-station) and $K$ receivers (users).
At each channel use ({\it slot}) the transmitter sends one symbol (packet) from the input alphabet $\mathcal{X}$, assumed to be a discrete finite set.
A packet either reaches a receiver error-free or it is erased (lost).
Erasures are assumed to be independent and identically distributed (iid) across channel uses, that is, the channel is memoryless.
A fixed number of bits $\lambda_k \log_2(|\mathcal{X}|)$ arrives at the transmitter every $T_k$ slots and must be delivered within the next arrival to receiver $k\in[1:K]$.
We refer to $T_k$ as the {\it hard deadline} for receiver $k\in[1:K]$.

We let $T := \mathsf{LCM}(T_1, \ldots, T_K)$, where $\mathsf{LCM}$ stands for least common multiple.
A group of $T$ slots is referred to as a {\it frame} and represents the time window
over which the transmitter needs to make joint scheduling decisions. 
For receiver $k\in[1:K]$, each frame 
is composed of $T/T_k$ {\it sub-frames}. 
 {\it Blocks} are the groups of time slots between consecutive deadlines considering both channels. Sub-frames and blocks will be illustrated next; blocks are of importance as we will make scheduling decisions block by block.

\begin{example}
\label{ex:2*T1=T2}
In Fig.~\ref{fig:network}, (a) illustrates the described communication system for the case of $K=2$ receivers, (b) and (c) illustrate two possible erasure pattern realization for a frame of length $T=6$ slots for the case of $2T_1 = T_2$ and $3T_1 = 2T_2$, respectively. 
Erased slots are shown with a hatch pattern,  erasure-free slots with 
white color.
$\lambda_1$  (resp. $\lambda_2$) is the amount of data 
intended for the first (resp. second) receiver and is generated every $T_1=3$ (resp. $T_2=6$) slots in (a) and every $T_1=2$ (resp. $T_2=3$) slots in (b) .

Slots forming a sub-frame are shown with dotted ovals while a group of slots shown with horizontal parenthesis form a block. Notice here that the sub-frames are defined for each user but blocks are common to both users.

\begin{figure}
\centering
\subfloat[]{\label{main:c}\includegraphics[width=70mm]{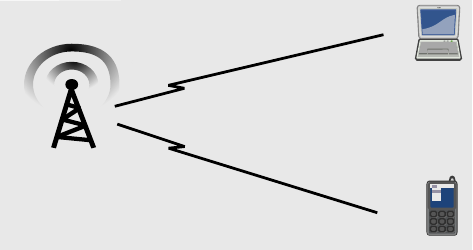}}

\begin{minipage}{.5\linewidth}
\centering
\subfloat[]{\label{main:a}\includegraphics[width=40mm]{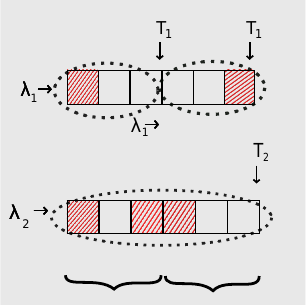}}
\end{minipage}%
\begin{minipage}{.5\linewidth}
\centering
\subfloat[]{\label{main:b}\includegraphics[width=40mm]{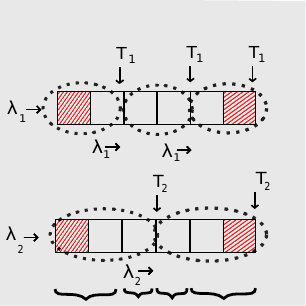}}
\end{minipage}\par\medskip

\vspace{-10pt}
\caption{(a) Illustrates a down-link broadcast network with one base station and two receivers, (b) illustrates the erasure broadcast channel where $2T_1=T_2 $, (c) illustrates the erasure broadcast channel where $3T_1=2T_2 $. Dotted ovals indicate sub-frames, while horizontal parentheses denote the blocks.}
\label{fig:network}
\vspace{-10pt}
\end{figure}

\end{example}

Let $\mathbf{E}_t \in [0:1]^{K\times 1}$
denote the (random) \textit{erasure pattern} in slot $t$,
where $[\mathbf{E}_t]_k=0$ means that the transmitted packet in slot $t\in[1:T]$ has not been received by receiver $k\in[1:K]$, and $[\mathbf{E}_t]_k=1$ otherwise.
Denote the erasure probabilities as
$\varepsilon_v := \Pr [\mathbf{E}_t = v]$, where we note that 
 $\sum_{v \in [0:1]^{K\times 1}} \varepsilon_v = 1$. 
Let $\mathbf{E} := [\mathbf{E}_1,\ldots,\mathbf{E}_T] \in [0:1]^{K\times T}$ be the information available at the transmitter in order to make scheduling decisions for a frame. 

The {\it erasure pattern probability} is
\begin{eqnarray}
\Pr[\mathbf{E} = \mathbf{e}] = \!\!\!\!
\prod_{v \in [0:1]^{K\times 1}} \left(\varepsilon_v\right)^{n_v(\mathbf{e})},
\
n_v(\mathbf{e}) := \!\!\!\!
\sum_{t\in[1:T]} 1_{\{\mathbf{e}_t = v\}} 
\label{eq:erasure pattern probability}
\end{eqnarray}
where $1_{\mathcal{A}}$ is the indicator function, which is one if the condition in $\mathcal{A}$ is true and zero otherwise and where $\mathbf{e} = [\mathbf{e}_1,\ldots,\mathbf{e}_T]\in [0:1]^{K\times T}$.

\begin{subequations}
A {\it scheduling policy} is a mapping 
\begin{align}
\Pi &: [0:1]^{K\times T} \to [0,1]^{K\times T} 
\notag
\\  &: \mathbf{e} \to \Pi(\mathbf{e}) = [\Pi_1(\mathbf{e}), \ldots, \Pi_T(\mathbf{e})],
\end{align}
where $[\Pi_t (\mathbf{e})]_k \in[0,1]$ is the amount of data, as a fraction of the slot capacity $\log_2(|\mathcal{X}|)$, sent to receiver $k\in[1:K]$ in slot $t\in[1:T]$ when the erasure pattern is $\mathbf{e} \in [0:1]^{K\times T}$.
A feasible policy 
must satisfy 
$\sum_{k\in[1:K]} [\Pi_t(\mathbf{e})]_k \leq 1,  \ \forall t\in[1:T]$, 
as it is not possible to transmit above the slot capacity.
\label{eq:policy def}
\end{subequations}
We say that {\it deadlines of all $K$ receivers are met} by the policy $\Pi$ for erasure pattern $\mathbf{e} = [\mathbf{e}_1,\ldots,\mathbf{e}_T]\in [0:1]^{K\times T}$ if and only if the arrival rates are within the following set
\begin{align}
\mathcal{E}_{\Pi}(\mathbf{e}) := \Big\{
  &\lambda_k \leq \sum_{t\in[1:T_k]} 
  [\mathbf{e}_{(m_k-1)T_k+t}]_k [\Pi_{(m_k-1)T_k+t}(\mathbf{e})]_k, 
\notag
\\&\forall m_k\in\left[1:T/T_k\right], \forall k\in[1:K] 
\Big\},
\label{eq:condition for no outage}
\end{align}
that is, each receiver $k\in[1:K]$ must be able to receive $\lambda_k$ packets within $T_k$ slots in every one of the $T/T_k$ sub-frames.

The objective is to find the policy in~\eqref{eq:condition for no outage} that minimizes the {\it global deadline outage probability}, defined as
\begin{align}
P_\text{out}(\lambda_1, \ldots, \lambda_K, T_1, \ldots, T_K) := 1 - \max_{\Pi}\Pr[\mathcal{E}_{\Pi}(\mathbf{E})],
\label{eq:pout def}
\end{align}
where the event $\mathcal{E}_{\Pi}(\cdot)$ is defined in~\eqref{eq:condition for no outage} as a function of the arrival rates $(\lambda_1, \ldots, \lambda_K)$ and the hard deadlines $(T_1, \ldots, T_K)$.

\begin{remark} \label{eq:pout use}
The ability to design rate control policies is a critical advantage of having a closed form expression for the deadline outage probability: the set of arrival rates $(\lambda_1, \ldots, \lambda_K)$ can be determined for a given desired set of deadlines $(T_1, \ldots, T_K)$ and a tolerable global outage probability $p$ by solving $P_\text{out}=p$ for $P_\text{out}$ given in~\eqref{eq:pout def}. 
Similarly, the set of deadlines by which packets can reach their destination can be predicted for a given fixed set of arrival rates and a deadline outage probability.
\end{remark}

This paper will deal with $T_2=NT_1$ and $K=2$ users, but we note that generalizations are possible to the general $MT_1=NT_2$. We do not present this here as the notation becomes quite involved.

Let $k$ denote the block index, $k \in [1:N]$
\begin{align}
N_{k,v}(\mathbf{E}) :=  
\sum_{t\in[1:T_1]} 1_{ \{ \mathbf{E}_{(k-1)T_1+t} = v \} }, 
\end{align}
be the random variable denoting the number of slots with erasure pattern $v\in [0:1]^{2\times 1}$. 
For notational convenience, we omit the dependence of $N_{k,v}$ on the random erasure pattern $\mathbf{E}$.

\section{A Greedy Scheduling Policy and its Optimality}
\label{sec:greedy}
{To obtain the scheduling policy that minimizes the global outage probability in \eqref{eq:pout def} (the scheduling policy that does so will be called ``optimal'' but need not be unique), we fix the erasure pattern ${\bf e}$, which in turn allows us to characterize a cut-set outer bound on the pairs of arrival rates $(\lambda_1, \lambda_2)$ that can be supported by that erasure pattern (in order to meet the hard deadlines). We then define a scheduling policy that achieves that outer bound for each possible erasure pattern, and hence minimizes the probability of outage, {\em i.e.,} the probability that not all the packets generated in the beginning of each frame can be served by the hard deadlines. }

\subsection{Outer Bound}

It is intuitive to see that an optimal policy has the following characteristics. 
In each block $k\in [1:N]$ of length $T_1$ slots:
i)   $N_{k,00}$ slots are useless;
ii)  all the $N_{k,10}$ slots must be allocated to receiver~1 (as they are useless to receiver~2)
and similarly the $N_{k,01}$ slots to receiver~2; 
iii) the $N_{k,11}$ slots must be split among the two receivers.
This intuition is confirmed by finding an optimal policy that reaches the cut-set bound -- a well-known outer bound on the channel capacity of networks in network information theory \cite{ElGamalKim:book} --  which we detail next for the channel in Fig.~\ref{fig:network} (b).

\begin{figure}
\begin{center}
\vspace{-2.5cm}
\noindent\includegraphics[width=70mm]{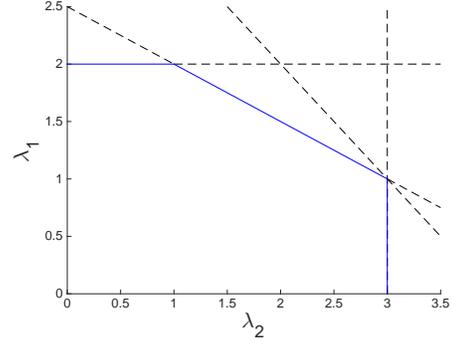}
\vspace{-60pt}
\caption{Cut-set upper bound for the erasure pattern shown in Fig.~\ref{fig:network}.}
\vspace{-20pt}
\label{region}
\end{center}
\end{figure}

\begin{subequations}
To describe the cut-set outer bound for a given erasure pattern ${\bf E} = {\bf e}$, let $n_{k,v}$,  $v\in [0:1]^{2\times 1}, k\in [1:N]$ denote the realizations of random variables $N_{k,v}$ corresponding to the realized erasure pattern ${\bf e}$. 
 For each block $k\in [1:N]$ define
\begin{align}
a_k &:= n_{k,10}+n_{k,11}, \\
b_k &:= n_{k,01}+n_{k,11}, \\
c_k &:= n_{k,10}+n_{k,01}+n_{k,11},
\end{align}
where $a_k$ is the number of slots available for transmission to receiver~1 without considering receiver~2,
 $b_k$ is the number of slots available for transmission to receiver~2 without considering receiver~1, 
and finally $c_k$ is the number of slots available for transmission to receivers~1 and~2 as if a ``super-receiver'' had the packets received by both receivers.
\end{subequations}
 
The (instantaneous, for a given erasure pattern realization) cut-set upper bound for the channel in Fig.~\ref{fig:network} ($T_2=2T_1$) can be characterized as follows.
Since the transmission to receiver~2 occupies the whole frame, which is made of $N=2$ blocks, the optimal scheduler decides to send $\zeta\lambda_2$ packets to receiver~2 during the first block and the remaining $(1-\zeta)\lambda_2$ during the second block, for some optimized ``rate split'' parameter $\zeta \in [0,1]$.
Receiver~1 is not in outage if $\lambda_1\leq a_1$ and $\lambda_1\leq a_2$;
receiver~2 is not in outage if
$\lambda_2\leq b_1+b_2$;
the combination of both receivers is not in outage if $\lambda_1+\zeta\lambda_2\leq c_1$ and $\lambda_1+(1-\zeta)\lambda_2\leq c_2$ for some $\zeta \in [0,1]$.
\begin{subequations}
The cut-set upper bound can thus be expressed as
\begin{align}
&
\bigcup_{\zeta \in [0,1]}
\begin{cases}
 \lambda_1\leq a_1,   
&\lambda_1\leq a_2,  \\
 \zeta\lambda_2\leq b_1, 
&(1-\zeta)\lambda_2\leq b_2, \\
 \lambda_1+\zeta\lambda_2\leq c_1, 
&\lambda_1+(1-\zeta)\lambda_2\leq c_2, \\
\end{cases}
\label{eq:cuset 1}
\\&=
\begin{cases}
\lambda_1            \leq \min(a_1,a_2), \\
           \lambda_2 \leq v_0, & \hspace*{-2mm} v_0:= b_1+b_2, \\
 \lambda_1+\lambda_2 \leq v_1, & \hspace*{-2mm} v_1:= \min(b_1+c_2,c_1+b_2), \\
2\lambda_1+\lambda_2 \leq v_2, & \hspace*{-2mm} v_2:= c_1+ c_2, \\
 \\ 
\end{cases}
\label{eq:cuset 2}
\\&=
\begin{cases}
\lambda_1 \leq \min(a_1,a_2), \\ 
\lambda_2 \leq v_0 - (\lambda_1-v_1+v_0)^+ - (\lambda_1-v_2+v_1)^+ \\
\phantom{\lambda_2} = v_0 - (\lambda_1-N_{1,10})^+ - (\lambda_1-N_{2,10})^+. 
\label{eq:cuset 4}
\end{cases}
\end{align}
because $v_1-v_0 = \min(c_1-b_1,c_2-b_2) = \min(n_{1,10}, n_{2,10})$
and     $v_2-v_1 = \max(c_1-b_1,c_2-b_2) = \max(n_{1,10}, n_{2,10})$, and
where $(x)^+ := \max(0,x)$. Such a region is illustrated in Fig.~\ref{region} for the erasure pattern shown in Fig.~\ref{fig:network} (a).
\end{subequations}

By extending to a generic integer $N$ (from $T_2=NT_1$) the previous reasoning 
we see that the cut-set bound reads
\begin{align}
\begin{cases}
\lambda_1\leq \min_{k\in [1:N]}(a_k),\\
\lambda_2 \leq v_k - k \lambda_1
= v_0 -\sum_{k=1}^{k}(\lambda_1 - v_{k} + v_{k-1})^+,\\ 
v_k := \min_{\mathcal{S}\subseteq [1:N] : |\mathcal{S}|=k}(\sum_{k\in\mathcal{S}} c_k+\sum_{k \not\in \mathcal{S}} b_k).\\
\end{cases}
\label{eq: cutset N*T1=T2}
\end{align}

or equivalently

 \begin{align} 
\mathcal{E}_{\Pi}(\mathbf{e}) := 
\begin{cases}
\lambda_1\leq \min_{k\in [1:N]}(a_k),\\
\lambda_2 \leq  v_0 - \sum_{k=1}^{N}(\lambda_1-N_{k,10})^+.
\end{cases}
\label{eq: final cutse}
\end{align}
where $v_0=\sum_{k=1}^{N} b_k$.

\begin{remark}
The generalization of~\eqref{eq: cutset N*T1=T2} to the case $MT_1 = N T_2$ follows by the same reasoning;
in this case the outer bound contains bounds of the form $k_1 \lambda_1 + k_2 \lambda_2 \leq v_{k_1,k_2}$
for some real-value $v_{k_1,k_2}$ and integers $(k_1,k_2)\in[1:T/T_1]\times[1:T/T_2]$.
Generalization to more than two receivers also follows easily
and the cut-set bound contains bounds on integer-valued linear combinations of the arrival rates.
\end{remark}

\subsection{\label{sec:greedy_policy} Greedy Policy}

We now demonstrate a scheduling policy
that can achieve the cut-set bound in~\eqref{eq: cutset N*T1=T2}, 
which will be first described for an example and then in a general form.
\begin{example}
\label{ex:2*T1=T2:out}
Consider the communication system shown in Fig.~\ref{fig:network} (b). In the first block, slot 3 is allocated to user 1 since it is useless to user 2.   The only real question is how to allocate slots which are simultaneously available to both users  (slot 2). Our ``greedy'' policy prioritizes the user with the sooner deadline, {\em i.e.,} the more urgent packet. In this example, the first user's packet has priority to be sent over slot 2 since its deadline in this time slot is earlier than the second user packets. If the capacity of this slot is completely used by this user, the second user cannot use that slot. If this slot is not completely occupied by user 1, the remaining capacity can be allocated to user 2. With the same reasoning, in the second block, slot 4 and 6 is allocated to user 1 and 2, respectively. Since the deadline of both packets are the same in time-slot 5, none of their packets has priority to be sent over this time-slot. 
\end{example}

{\bf Greedy policy definition:} Given a priori and full knowledge of the erasure pattern of each block $k\in [1:N]$, the greedy policy $\Pi$ in each block first allocates all the $N_{k,10}$ slots to receiver~1's packets, and all the $N_{k,01}$ slots to receiver~2's packets. After allocating all $N_{k,10}$ and $N_{k,01}$ slots, the $N_{k,11}$ slots in the same block should be allocated to the remaining packets of each user (if any) as follows: priority is given to the user with the earlier deadline until that user's packets are all sent; then they are allocated to the other user. If the capacity of such slots are completely used by the first receiver message, the second user will not be allocated any slots. Otherwise, the remaining capacity can be allocated to the second receiver message.
This block by block scheduling continues until a scheduling decision is made over a frame of length $T$. Note that scheduling decisions occur block by block and that the base-station  {\it only needs  knowledge of the erasure pattern of that block, knowledge of the erasures in the full frame is not needed!} However, in order to calculate the probability of outage, we need to keep track of whether the deadlines were met within the entire frame, consisting of multiple blocks in general.

\subsection{Optimality of the Greedy Policy}
\begin{theorem}
The greedy policy is optimal.
\end{theorem}

\begin{IEEEproof}  
Our central proof idea is to show that for a given erasure pattern ${\bf e}$, the greedy policy schedules packets so as to achieve the cut-set upper bound for that erasure pattern. Thus, the greedy policy supports the largest set of arrival rates $(\lambda_1, \lambda_2)$ for a given erasure pattern, {\em i.e.} minimizes the probability of outage given an erasure pattern. Summing over the probability of each erasure pattern yields the minimal probability of outage. 

To prove that for a given erasure pattern, the greedy policy minimizes the probability of outage, as mentioned before, our greedy policy first allocates the $N_{k,10}$ and $N_{k,01}$ (this is why we need full CSI) and then allocates $N_{k,11}$ slots greedily for the remaining packet (if any). Since $N_{k,10}$ and $N_{k,01}$ are usable by only one user, it is obvious that allocating such slots to packets in the first step of our greedy policy is optimal. Here we focus on optimality proof of using the $N_{k,11}$ slots greedily ({\em i.e.,} earliest deadline first until that users' packets are fully satisfied).

Suppose now that there exists an optimal scheduler that does not prioritize the user with most urgent message, and  that this optimal scheduler can find a way to send data before their deadlines while the greedy policy cannot.  In  the optimal non-greedy scheduler, starting from first simultaneous erasure-free time slot, sending data in such slots and moving forward, we will reach a simultaneous erasure-free slot, $t_i$, that is not assigned to the user with closer deadline (user 1 in this paper) while this user has still some data to receive. So its remaining data should be sent in one of the next $t_j$ simultaneous erasure-free slots (such slot exists before the deadline of user 1 since the optimal scheduler is able to send packets and all $N_{k,10}$ and $N_{k,01}$ slots have been occupied). 
This amount of data can also be sent in  the $t_i^{th}$ slot if it is not used by the second user. Even if it is occupied by the second user, we can send the second user packet in the $t_j^{th}$ slot since $t_j \leq T_2$.  Allocating the $t_i^{th}$ slot to the first user does not cause an outage and the greedy policy is also able to transfer the same packets as the optimal non-greedy scheduler. 
This contradicts our assumption that the greedy policy cannot send the packets prior to their deadlines. 
This reasoning is also applicable to the case $MT_1 = N T_2$.
\end{IEEEproof}

\section{Derivation of the Global Deadline Outage Probability}
\label{sec:global outage}

In this section, we derive the global deadline outage probability for the discrete memoryless broadcast erasure channel for two users where $T=T_2=N T_1$. {The derivation for $MT_2=NT_1$ will appear in a journal version of this work.}

The {\it deadline outage probability} can be calculated by plugging \eqref{eq: final cutse} into \eqref{eq:pout def}.  The greedy  policy satisfies the deadlines only for patterns for which  \eqref{eq: final cutse} holds.  Therefore, the {\it deadline outage probability}, can be calculated as follows. Let $\{N\}_k := \{n_{k,00}, n_{k,01}, n_{k,10}, n_{k,11} :  \sum_{m,n \in [0:1]} n_{k,mn} = \frac{T}{N}\}$. Then

\begin{align}
&P_\text{out}\left(\lambda_1, \lambda_2, \frac{T}{N}, T\right)=
 1-  \sum_{k=1}^N \sum_{\{N\}_k} \notag\\
&\prod_{j=1}^{N}{\frac{T}{N} \choose n_{j,00},n_{j,01},n_{j,10},n_{j,11}}  \Pr[\mathbf{E} = \mathbf{e}] \ \  1_{\{\mathcal{E}_{\pi}(\mathbf{e})\}}
\label{sec:pout closed form}
\end{align}
where in \eqref{sec:pout closed form} we see the notation for the multi-nomial coefficient and
$\Pr[\mathbf{E} = \mathbf{e}]$ can be obtained by \eqref{eq:erasure pattern probability}.

Note that for notational convenience, we omit the dependence of $N_{k,v}$ (or its instances $n_{k,v}$) on the erasure pattern $\mathbf{e}$.

\begin{figure}
\begin{center}
\noindent\includegraphics[width=80mm] {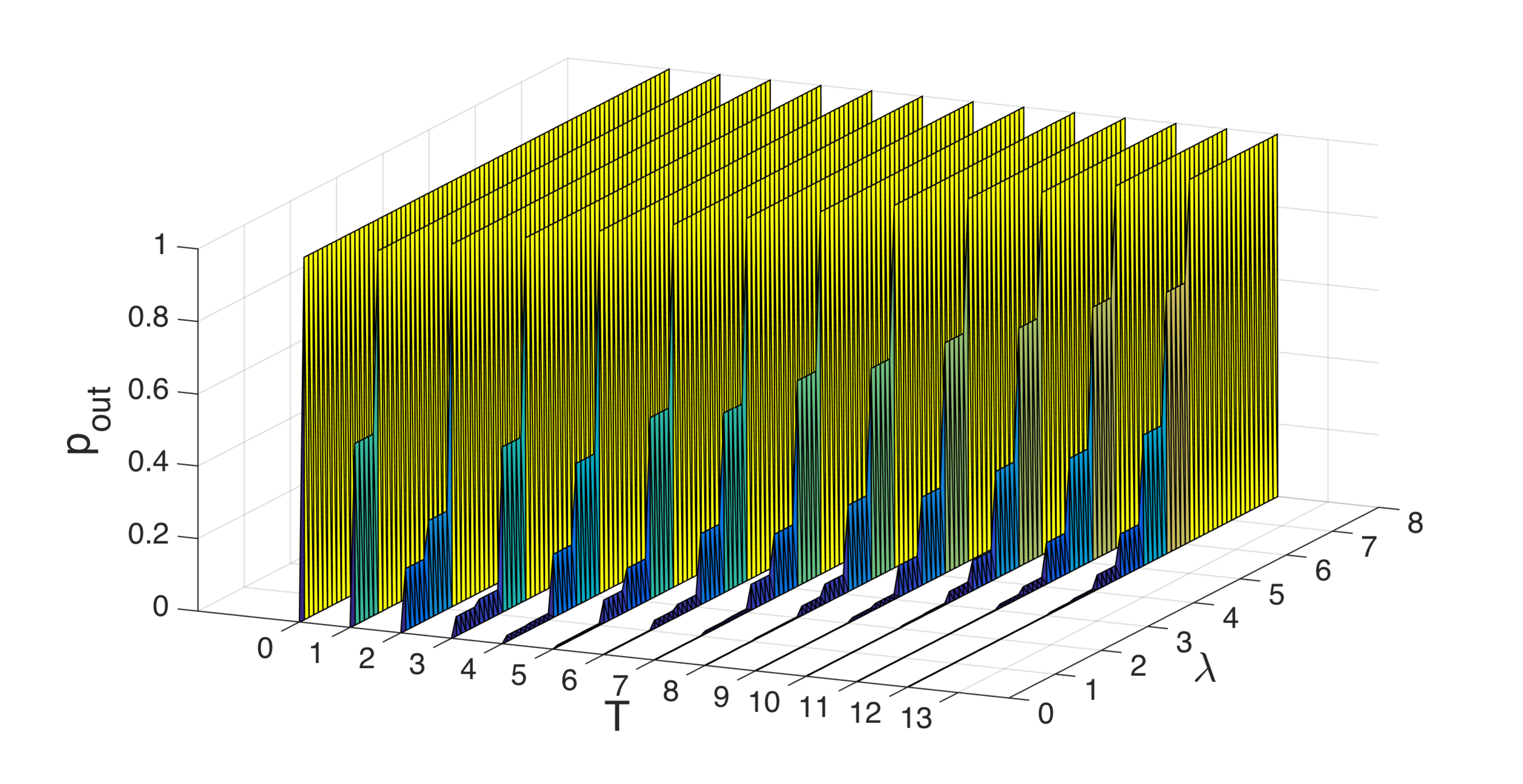}
\vspace{-10pt}
\caption{Probability of outage versus $\lambda$ and $T$ where $T=T_1=T_2,  \, \lambda=\lambda_1=\lambda_2, \, \epsilon_{01}= \epsilon_{10}=0.2, \epsilon_{11}= 0.5, \epsilon_{00}=0.1$.}
\label{fig:2}
\vspace{-10pt}
\end{center}
\end{figure}

\begin{figure}
\begin{center}
\noindent\includegraphics[width=70mm] {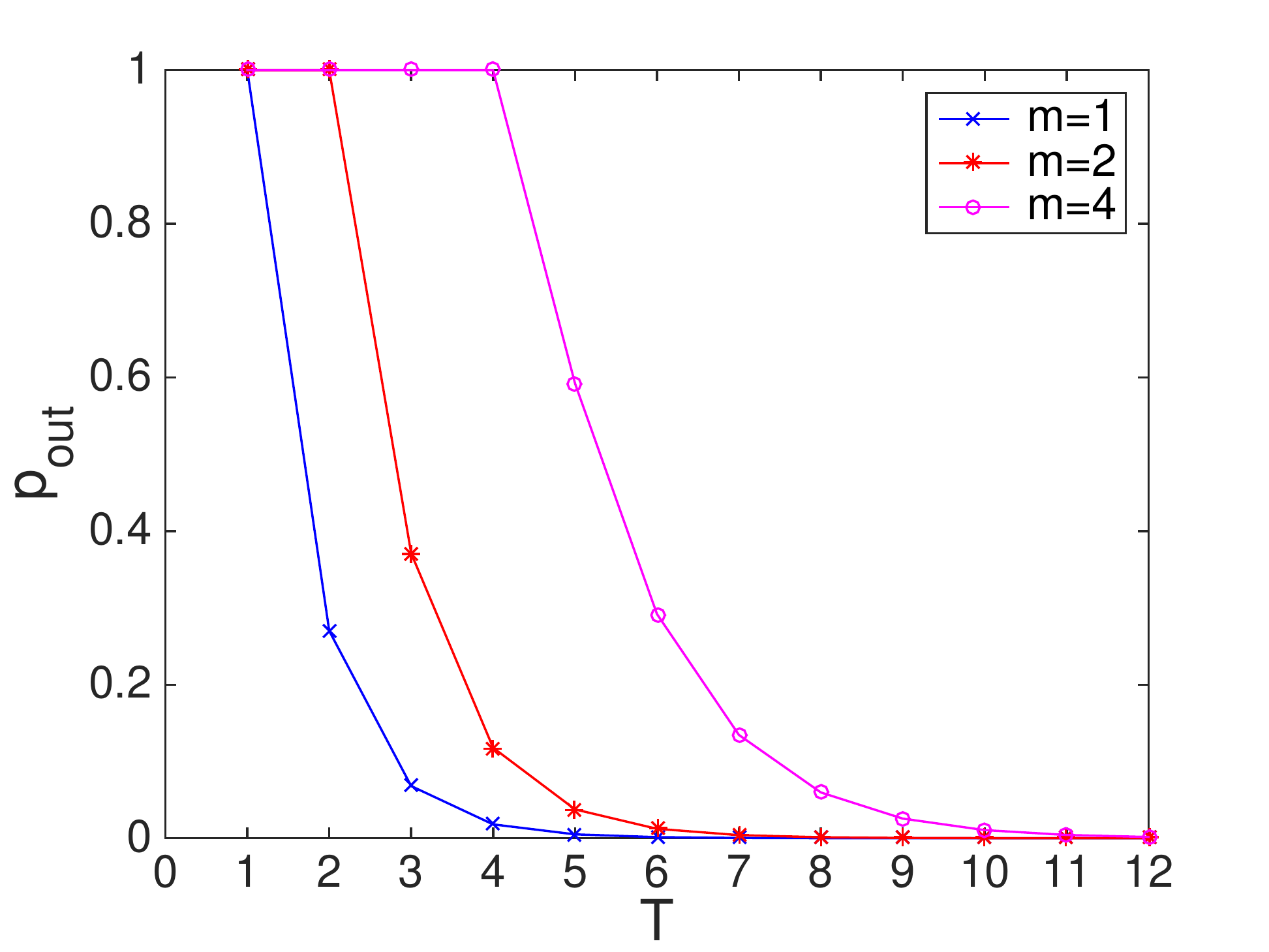}
\vspace{-10pt}
\caption{Probability of outage versus $T$ for some values of $m$ where $T=T_1= T_2$ and $\lambda_2=1, \lambda_1=m\lambda_2$, $\epsilon_{01}= \epsilon_{10}=0.2, \epsilon_{11}= 0.5, \epsilon_{00}=0.1$.}
\label{fig:3}
\vspace{-10pt}
\end{center}
\end{figure}

\begin{figure}
\begin{center}
\noindent\includegraphics[width=80mm] {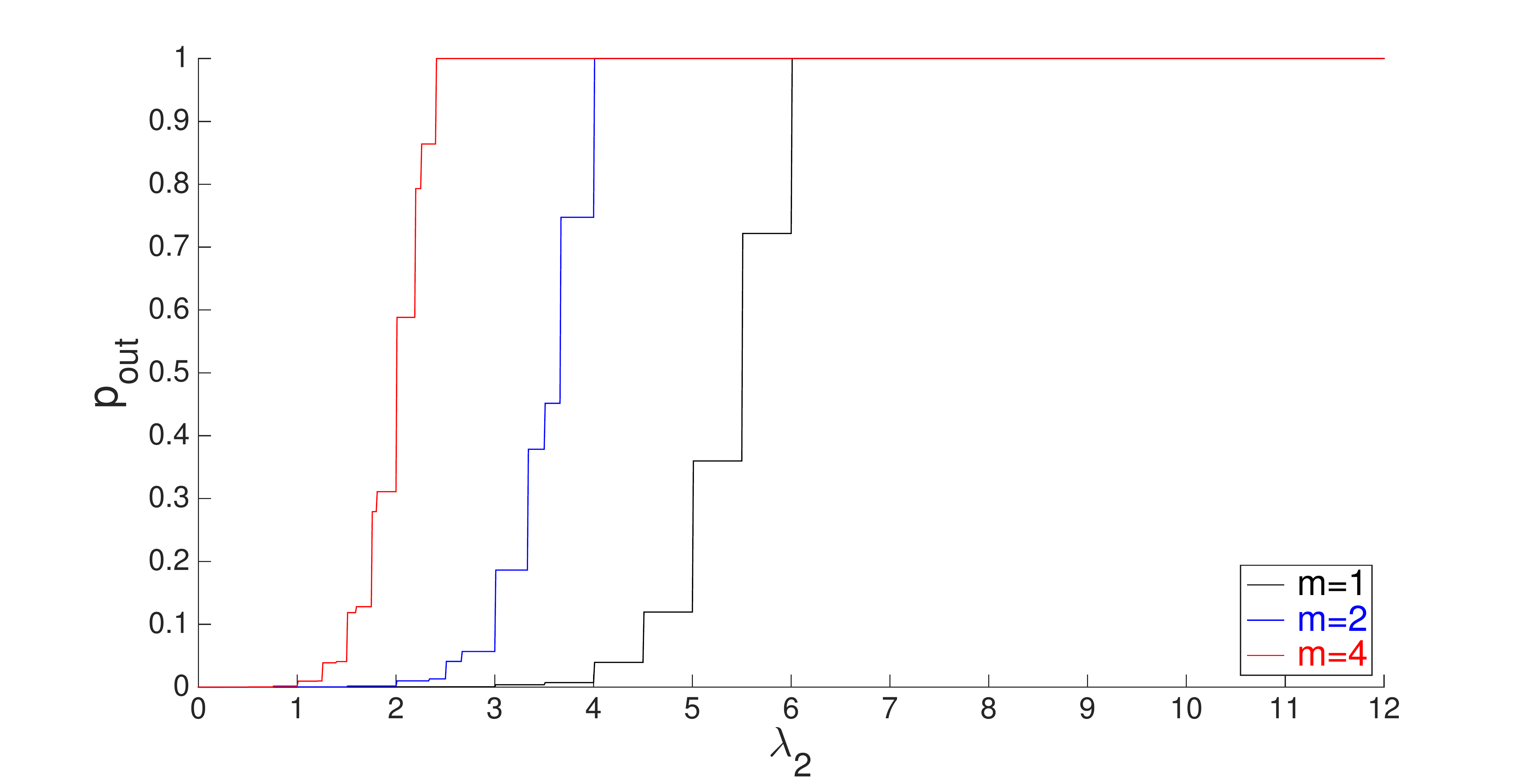}
\vspace{-10pt}
\caption{Probability of outage versus $\lambda_2$ for some values of $m$ where $T_1=T_2=12, \lambda_1=m\lambda_2,\epsilon_{01}= \epsilon_{10}=0.2, \epsilon_{11}= 0.5, \epsilon_{00}=0.1$.}
\label{fig:4}
\vspace{-10pt}
\end{center}
\end{figure}

\begin{figure}
\begin{center}
\noindent\includegraphics[width=80mm] {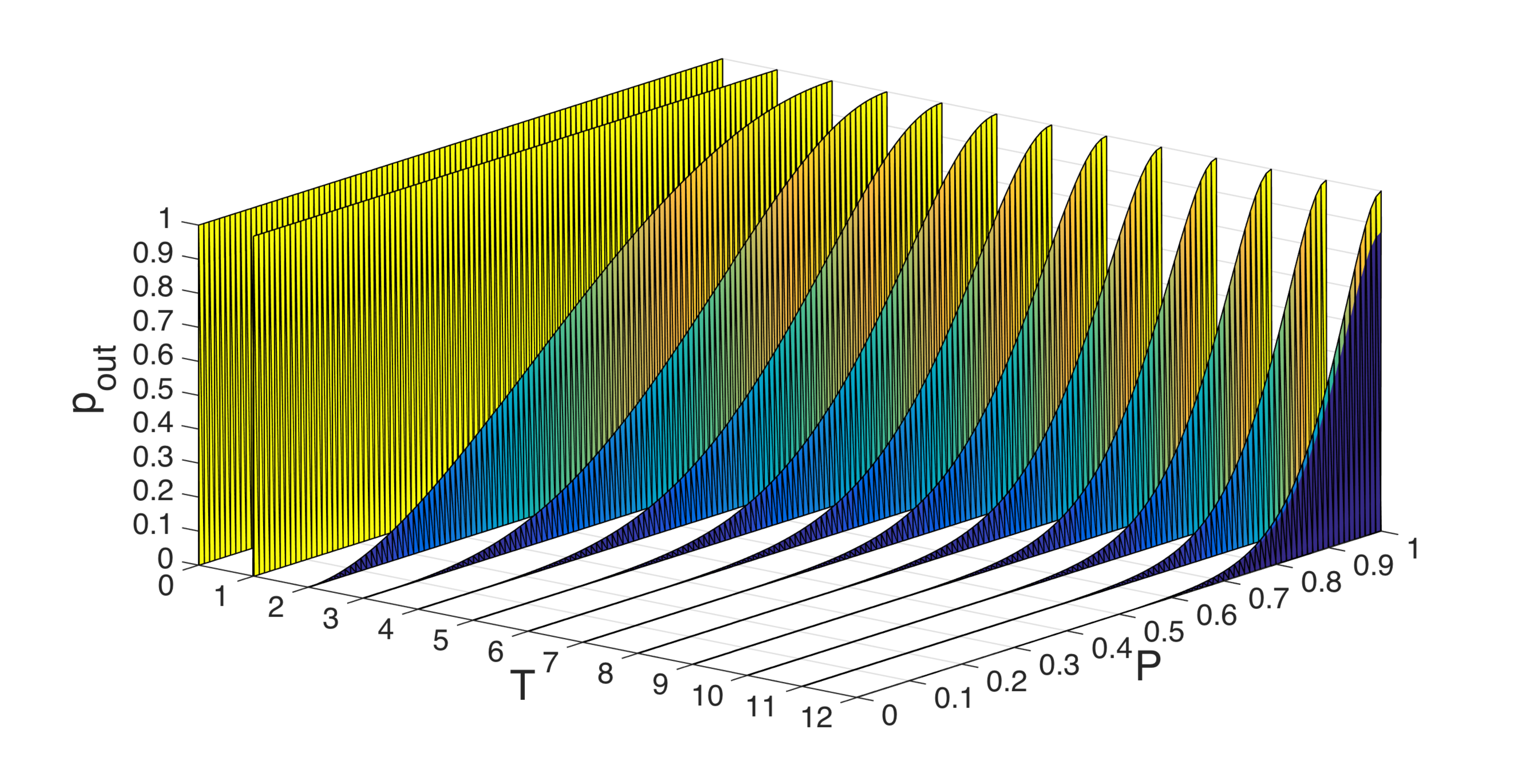}
\vspace{-10pt}
\caption{Probability of outage versus $T$ and $P$ where $T_1=T_2=12, \lambda_1=\lambda_2=1$ and $P$ is the erasure probability of each user.}
\label{fig:6}
\vspace{-10pt}
\end{center}
\end{figure}

Fig.~\ref{fig:2} illustrates the probability of outage for some values of $T$ and $\lambda$ where $T=T_1=T_2$ and $\lambda=\lambda_1=\lambda_2$. As expected, the probability of outage is zero when $\lambda=0$ for all values of $T$ and is one when $T=0$ for all values of $\lambda$. Moreover, for a fixed value of $T$, the probability of outage is increasing  in $\lambda$, and for a given value of $\lambda$ is decreasing in $T$.  The probability of outage jumps to the next value when either $\lambda_1$, $\lambda_2$  or $\lambda_1+\lambda_2$ reaches the next integer value. This is because the erasures are binary while $\lambda_1$ and $\lambda_2$  are non-negative and real-valued: the probability of outage remains constant as long as the free slot is not completely occupied by both users' traffic and jumps when it is completely used.

Fig.~\ref{fig:3} illustrates the probability of outage versus $T$ for some values of $m$ where $T_1=T_2=T$ and $\lambda_2=1$ and $\lambda_1=m\lambda_2$. It can be seen that by increasing $T$ the probability of outage goes to zero. For small $m$, the probability of outage tends to zero more quickly. 

Fig.~\ref{fig:4} illustrates the probability of outage for some values of $\lambda_2$ where $T_1=T_2=12$ and $\lambda_1=m\lambda_2$. It can be seen that by increasing $m$ the probability of outage goes to one more quickly.  The probability of outage is one when $\lambda_1+\lambda_2 > 12$ (as it would be impossible to serve more than 12 packets with a deadline of $T_1=T_2=12$).

Fig.~\ref{fig:6} illustrates the probability of outage versus $T$ and $P$ where $T=T_1=T_2=12, \lambda_1=\lambda_2=1$ and $P$ is the erasure probability of each user (erasures are assumed to be i.i.d. across time and users). Here we see the tradeoff between the channel parameter $P$, the deadline $T$ and the probability of outage. As expected, the larger the deadline for a fixed arrival rate, the smaller the probability of outage for a given $P$.

\section{Scheduling policy without future knowledge of channel states}
\label{sec:num}

In this section, we consider a more practical scenario where the future channel states are not known to the base station. As a first step, we assume that the current state is known to the base station, but the future states are not available. Then, we will consider the scenario that both current and future states are unavailable. 
These scheduling policies are heuristics at the moment -- proving optimality is left for future work.

We first consider that the scheduler (base station)  has the knowledge of the current channel state only. In particular, when we transmit at time slot $t \in [1:T]$, the scheduler knows the channel state information ${\bf E}_t$, but it does not know the channel states ${\bf E}_\tau$ s.t. $\tau > t$. For the future channel states, the scheduler only knows the erasure probabilities. Next, we discuss how a greedy algorithm is designed when the deadlines of both users is the same; {\em i.e.,} $T_1 = T_2 = T$. Our algorithm is summarized in Algorithm~\ref{alg:only_one_slot_known}. 

Let us consider that at each time slot $t$, the number of packets that are still supposed to be transmitted is $\lambda_1^{\text{rem}}$ and $\lambda_2^{\text{rem}} $ for users 1 and 2, and the remaining time  to schedule packets are $T^{\text{rem}}$.  We can express the global deadline outage probability in (\ref{eq:pout def}) as $P_{\text{out}}(\lambda_1^{\text{rem}}, \lambda_2^{\text{rem}}, T^{\text{rem}})$ for the packets after (and including) slot $t$ by abusing the notation (note that we abbreviate $P_{\text{out}}(\lambda_1^{\text{rem}}, \lambda_2^{\text{rem}}, T^{\text{rem}}, T^{\text{rem}} ) $ to $P_{\text{out}}(\lambda_1^{\text{rem}}, \lambda_2^{\text{rem}}, T^{\text{rem}})$).

      \begin{algorithm}[H]
        \caption{\label{alg:only_one_slot_known} Scheduling algorithm for hard deadlines with knowledge of the current channel state}
        \begin{algorithmic}
          	\State Init: $\lambda_1^{\text{rem}} = \lambda_1$, $\lambda_2^{\text{rem}} = \lambda_2$, $t=1$, $T^{\text{rem}} = T$
			\While{($T^{\text{rem}} > 0$)}
			\State $P_{\text{out}}^{\text{Xmit 1}} = P_{\text{out}}(\lambda_1^{\text{rem}} - 1, \lambda_2^{\text{rem}}, T^{\text{rem}}-1)$
			\State $P_{\text{out}}^{\text{Xmit 2}} = P_{\text{out}}(\lambda_1^{\text{rem}}, \lambda_2^{\text{rem}} - 1, T^{\text{rem}}-1)$
			\If{$\{\lambda_1^{\text{rem}}\} > 0$ AND $\{  {[\bf E}_t]_1 = 1$ AND ( $[{\bf E}_t]_2 = 0  $ OR $ (P_{\text{out}}^{\text{Xmit 1}} < P_{\text{out}}^{\text{Xmit 2}}) \}$}
			\State Transmit a packet to 1 
			\State $\lambda_1^{\text{rem}} = \max\{0,\lambda_1^{\text{rem}}-1\}$
			\ElsIf{ ${[\bf E}_t]_2 = 1$ AND $\lambda_2^{\text{rem}} > 0$}
			\State Transmit a packet to 2 
			\State $\lambda_2^{\text{rem}} = \max\{0,\lambda_2^{\text{rem}}-1\}$
			\EndIf
			\State $T^{\text{rem}} = T^{\text{rem}} - 1$
			\State $t = t + 1$
			\EndWhile
        \end{algorithmic}
      \end{algorithm}

If a packet is successfully transmitted from user 1 at time slot $t$, then the outage probability after time slot $t$ will be $P_{\text{out}}^{\text{Xmit 1}} = P_{\text{out}}(\lambda_1^{\text{rem}} - 1, \lambda_2^{\text{rem}}, T^{\text{rem}}-1)$, because (i) $\lambda_1^{\text{rem}} - 1$ packets will remain to be scheduled after transmitting a packet from user 1, (ii) the number packets that should be scheduled from user 2 will not change, {\em i.e.,} $\lambda_2^{\text{rem}}$ will not change, and (iii) the remaining time to schedule the packets will reduce to $T^{\text{rem}}-1)$. Similarly, we can define the outage probability as $P_{\text{out}}^{\text{Xmit 2}} = P_{\text{out}}(\lambda_1^{\text{rem}}, \lambda_2^{\text{rem}} - 1, T^{\text{rem}}-1)$ after a packet from user 2 is scheduled at time slot $t$. 

Our scheduling policy examines the channel state at slot $t$. If $  {[\bf E}_t]_1 = [{\bf E}_t]_2 = 0$, we do not schedule any packets at slot $t$. If $  {[\bf E}_t]_1 = 1, [{\bf E}_t]_2 = 0  $ and $  {[\bf E}_t]_1 = 0, [{\bf E}_t]_2 = 1  $, we schedule users 1 and 2, respectively. On the other hand, if $ {[\bf E}_t]_1 = [{\bf E}_t]_2 = 1$, we compare $P_{\text{out}}^{\text{Xmit 1}}$ and $P_{\text{out}}^{\text{Xmit 2}}$, and schedule a packet for user 1 if $ (P_{\text{out}}^{\text{Xmit 1}} < P_{\text{out}}^{\text{Xmit 2}})$, otherwise schedule a packet for user 2. 

\begin{figure}
\begin{center}
\vspace{-5pt}
\noindent\includegraphics[width=80mm]{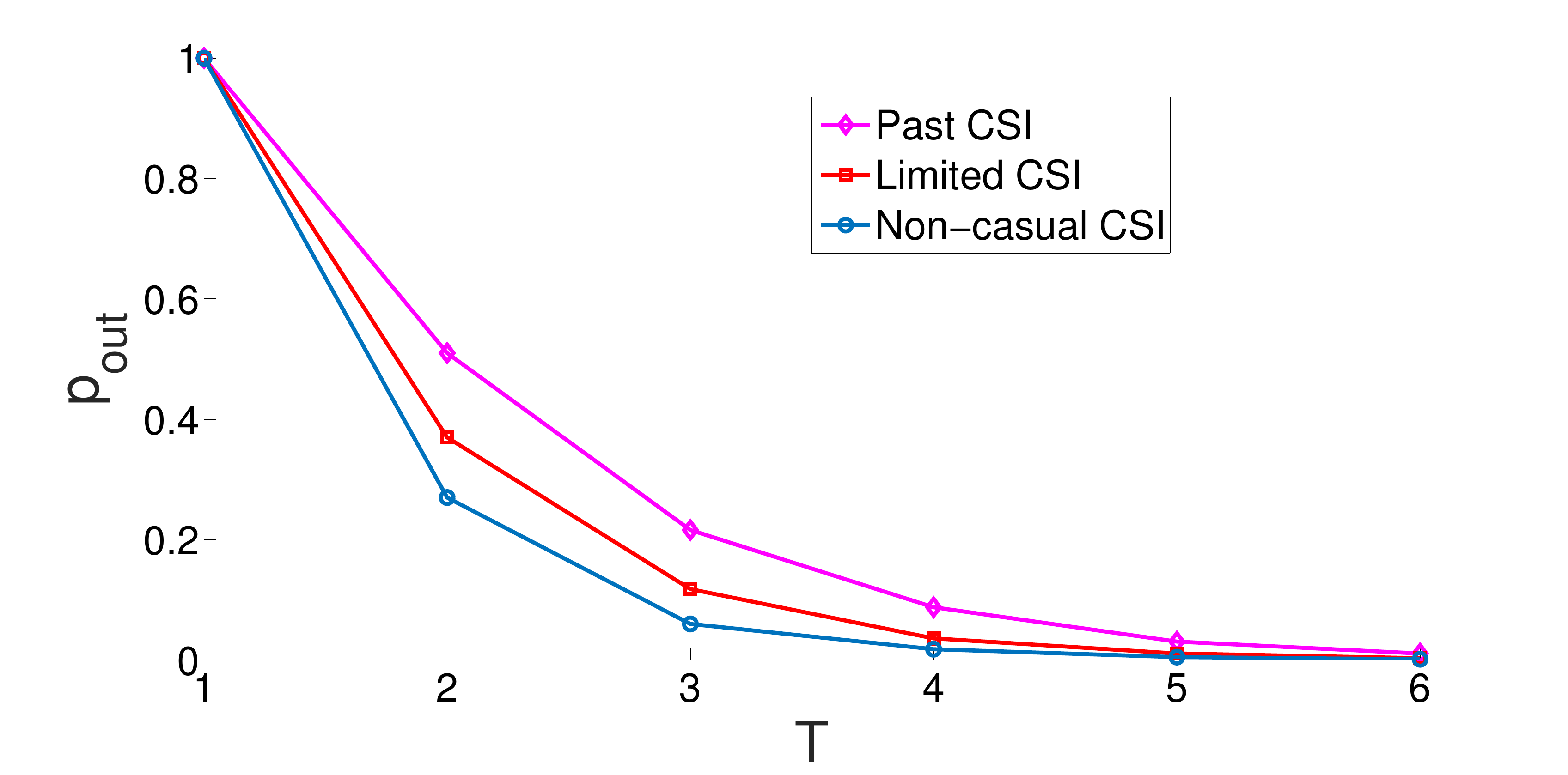}
\vspace{-10pt}
\caption{$P_\text{out}$ versus $T$, when $\lambda_1=\lambda_2=1$ and $\epsilon_{01}= \epsilon_{10}=0.2, \epsilon_{11}= 0.5, \epsilon_{00}=0.1$.
  (i) Non-casual CSI: The greedy algorithm in Section~\ref{sec:greedy_policy}. 
 (ii) Current CSI: Algorithm~\ref{alg:only_one_slot_known}. 
(iii) Past CSI.}
\label{fig:all_algs}
\vspace{-20pt}
\end{center}
\end{figure}

We evaluated Algorithm~\ref{alg:only_one_slot_known} when $\lambda_1=\lambda_2=1$ and $\epsilon_{01}= \epsilon_{10}=0.2, \epsilon_{11}= 0.5, \epsilon_{00}=0.1$. Fig.~\ref{fig:all_algs} shows the outage probability versus deadline. As expected, Algorithm~\ref{alg:only_one_slot_known} performs worse than non-casual CSI (the greedy algorithm in Section~\ref{sec:greedy_policy}), but the outage probability delay curves follow the same decay pattern. The proof of optimality for this algorithm (optimal in the sens of minimizing the globale outage probability under these limited CSI constraints) is left for future work. 

We now assume that in slot $t$, the channel states of the past slots up to time $t_1$ are available (perfectly or partially), but that the current and future states are unavailable. In this setup, we make scheduling decisions at the start of the slot and receive feedback on successful (or not) transmission at the end of the slot (we assumed no delayed feedback). In particular, at the start of the slot $t$, we calculate the outage probability when a packet from user 1 is transmitted as $P_{\text{out}}^{\text{Xmit 1}} =  P_{\text{out}}(\lambda_1^{\text{rem}} - 1, \lambda_2^{\text{rem}}, T^{\text{rem}}-1)(\epsilon_{10} + \epsilon_{11}) + P_{\text{out}}(\lambda_1^{\text{rem}} , \lambda_2^{\text{rem}}, T^{\text{rem}}-1)(\epsilon_{00} + \epsilon_{01})$. The first term, {\em i.e.,} $ P_{\text{out}}(\lambda_1^{\text{rem}} - 1, \lambda_2^{\text{rem}}, T^{\text{rem}}-1)(\epsilon_{10} + \epsilon_{11})$ corresponds to successful transmission scenario, while the second term, {\em i.e.,}  $P_{\text{out}}(\lambda_1^{\text{rem}} , \lambda_2^{\text{rem}}, T^{\text{rem}}-1)(\epsilon_{00} + \epsilon_{01})$ corresponds to the failure event. Our proposed scheduling policy compares outage probabilities, and if $P_{\text{out}}^{\text{Xmit 1}} < P_{\text{out}}^{\text{Xmit 2}}$, a packet is transmitted to user 1, otherwise to user 2. At the end of the slot, the scheduler receives perfect and immediate feedback from the users, and determines if packets are correctly delivered. Note that if a packet is not correctly delivered, the remaining packets; {\em i.e.,} $\lambda_1^{\text{rem}} $ and $\lambda_2^{\text{rem}} $ does not reduce. This means the same packet will be re-transmitted again. The global deadline outage probability with past CSI is shown in Fig.~\ref{fig:all_algs}.  Past CSI performs worse than non-casual CSI and limited CSI. This confirms the necessity of optimizing re-transmission mechanisms, which is left for the future work.

\section{Conclusion}
\label{sec:conclusion}

This paper studies the scheduling problem for transmitting periodically generated flows with hard-deadline constraints over an erasure broadcast channel. 
We propose a `greedy' scheduling policy (serving the user with the earliest deadline first) that is optimal in the sense that it minimizes the {\it global deadline outage probability} when the channel erasures are known ahead of transmission (only knowledge block by block is needed). We obtain a closed form expression for this global probability of outage, yielding a tradeoff between the arrival rates, the hard deadlines, and the reliability (probability of meeting those deadlines). 
Furthermore, two heuristics are proposed for more practical scenarios, where the channel state information is not known to the base station ahead of time. Future work includes extensions to per-user probabilities of outage, extensions to Gaussian broadcast channels, and extensions to the dual of this problem -- the uplink multiple-access channel.

\bibliography{referenc}{}
\bibliographystyle{plain}

\end{document}